\documentclass[10pt]{article}
\usepackage{amsmath}
\usepackage{amsfonts}
\usepackage{amssymb}
\usepackage{epsfig}
\usepackage{times}
\usepackage{oldgerm}
\newcommand{\be}{\begin{equation}}
\newcommand{\ee}{\end{equation}}
\newcommand{\ba}{\begin{array}}
\newcommand{\ea}{\end{array}}
\newcommand{\bea}{\begin{eqnarray}}
\newcommand{\eea}{\end{eqnarray}}

\newcommand{\rar}{\rightarrow}

\newcommand{\la}{\langle}
\newcommand{\ra}{\rangle}
\newcommand{\bs}{\boldsymbol}

\renewcommand{\l}{\newline\null}
\def\figskip{\vskip .5cm plus 3mm minus 2mm}
\def\hbar{\not{\hbox{\kern-2.3pt $h$}}}
\def\psl{\not{\hbox{\kern-2.3pt $p$}}}
\def\Psl{\not{\hbox{\kern-2.3pt $P$}}}
\def\ksl{\not{\hbox{\kern-2.3pt $k$}}}
\def\qsl{\not{\hbox{\kern-2.3pt $q$}}}
\begin{document}
\begin{titlepage}
%
July 1999 (revised February 2000)\hfill PAR-LPTHE 99/28
\vskip 5cm
{\baselineskip 17pt
\begin{center}
{\bf EXTENDING THE STANDARD MODEL:\break
AN UPPER BOUND FOR A NEUTRINO MASS
FROM THE RARE DECAY $\bs{K^+ \rightarrow \pi^+ \nu \bar\nu}$}
\end{center}
}
\vskip .5cm
\centerline{B. Machet
     \footnote[1]{Member of `Centre National de la Recherche Scientifique'}
     \footnote[2]{E-mail: machet@lpthe.jussieu.fr}
     }
\vskip 5mm
\centerline{{\em Laboratoire de Physique Th\'eorique et Hautes \'Energies}
     \footnote[3]{LPTHE tour 16\,/\,1$^{er}\!$ \'etage,
          Universit\'e P. et M. Curie, BP 126, 4 place Jussieu,
          F-75252 Paris Cedex 05 (France)}
}
\centerline{\em Universit\'es Pierre et Marie Curie (Paris 6) et Denis
Diderot (Paris 7)}
\centerline{\em Unit\'e associ\'ee au CNRS UMR 7589}
\vskip 1.5cm
{\bf Abstract:}  The standard model seeming at a loss to account for the
present experimental average rate for the rare decay
$K^+ \rightarrow \pi^+ \nu\bar\nu$, I tackle the question with the extension
of the Glashow-Salam-Weinberg model to an $SU(2)_L \times U(1)$ gauge theory
of $J=0$ mesons proposed in \cite{Machet1},
in which, in addition, the neutrinos are given Dirac masses from Yukawa
couplings to the Higgs boson. The latter triggers a new contribution to this
decay through flavor changing neutral currents that arise in the quartic term
of the symmetry breaking potential; it becomes sizeable for a neutrino mass
in the $MeV$ range;  the experimental upper
limit for the decay rate translates into an upper bound of $5.5\,MeV$ for
the mass of the neutrino, three times lower than present direct bounds.
\smallskip

\vfill
\null\hfil\epsffile{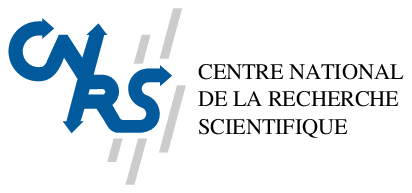}
\end{titlepage}
%
%
\section{Introduction}
\label{section:introduction}

The absence of flavor changing neutral currents at tree level is
one of the most severe criteria to select among the various
candidates which could generalize the Glashow-Salam-Weinberg model of
electroweak interactions. They are in general linked to higher
order corrections and their experimental quest is of high interest for the
pursuit of theoretical investigations and the search for hints of new
physics beyond the standard model.

Among the decays of particular interest involving flavor changing neutral
currents are
\begin{equation}
K^+ \longrightarrow \pi^+ \ell \bar\ell
\label{eq:reac}
\end{equation}
where the $\ell$'s stand for leptons which can be charged (electrons or
muons) or neutral (neutrinos).

One event compatible with a two-neutrinos
final state has recently been observed \cite{BNL787} corresponding to
a branching ratio
\begin{equation}
BR(K^+ \rightarrow \pi^+ \nu\bar\nu)_{exp} = 4.2
\begin{matrix}{+9.7}\\{-3.5}\end{matrix}\;10^{-10},
\label{eq:exBRnunu}
\end{equation}
while recent theoretical calculations \cite{Buras} find, for massless
neutrinos, an absolute upper bound
\begin{equation}
BR(K^+ \rightarrow \pi^+ \nu\bar\nu)_{th} < 1.22 \;10^{-10},
\label{eq:thBRnunu}
\end{equation}
and the authors of \cite{Buras} claim
``{\it a clear conflict with the Standard Model if $BR(K^+
\rightarrow \pi^+ \nu\bar\nu)$ should be measured at $2\;10^{-10}$}\ '',
that is less than one-half the experimental average (\ref{eq:exBRnunu}).

At the same time, we are  witnessing a dramatic change in our perception of
neutrinos since there seems to be more and more compelling evidence
\cite{Bilenky} that their flavor eigenstates oscillate \cite{Petcov}
during their travel across vacuum or matter, which is the best
model-independent sign of them being massive.

The question that can naturally be raised is whether massive neutrinos can
substantially increase the rate of the above decay in the standard framework,
specially through penguin-like diagrams where the Higgs boson couples to the
internal $W$ gauge boson or to the very massive top quark. The computations have
just been performed \cite {Perez} in the case of $K_L \rar \pi^0 \nu\bar\nu$
and showed that the influence of massive neutrinos is totally negligeable
when mass and flavour leptonic eigenstates coincide, and has a
relative upper bound of no more than $1/10$ when flavour mixing is allowed.

Rare semi-leptonic $K$ decays provide consequently a good testing ground
for physics beyond the standard model \cite{GrossmanNir}.

I investigate below the influence of neutrino masses on the decay
$K^+ \rightarrow \pi^+ \nu\bar\nu$ in the framework of the extension of
the Glashow-Salam-Weinberg model to an $SU(2)_L \times U(1)$ gauge theory of
$J=0$ mesons proposed in \cite{Machet1}.
 
It is built with a maximum compatibility with the standard model in the
quark-gauge sector
\footnote{Relying on this, the contributions involving
gauge bosons will be assumed to be of the same order of magnitude as when
computed in the standard model.}
: chiral and electroweak properties of quarks are included from the start
\footnote{The  $SU(2)_L \times U(1)$ electroweak group is embedded into the
larger
chiral $U(N)_L \times U(N)_R$ group ($N/2$ is the number of generations),
which is only possible when $N$ is even; this is to be compared with chiral
perturbation theory \cite{chpt}, which is always performed for an odd (three)
number of flavours; in the present framework, the relevant chiral breaking
appears instead to be the one of $SU(2)_L \times SU(2)_R$ into its diagonal
subgroup, the custodial $SU(2)$.
}
; it however differs in the Higgs-scalar
sector in that the three Goldstones of the broken symmetry(es) are now
related \cite{Machet2} through a scaling factor (see (\ref{eq:b}) below)
to pseudoscalar mesons and the Higgs boson is naturally
incorporated as one among the $J=0$ mesons, which all are considered to
transform like quark-antiquark composite fields.
The  $SU(2)_L \times U(1)$ electroweak group is embedded into the larger
chiral $U(N)_L \times U(N)_R$ group ($N/2$ is the number of generations),
which is only possible when $N$ is even; this is to be compared with chiral
perturbation theory \cite{chpt}, which is always performed for an odd (three)
number of flavours; in the present framework, the relevant chiral breaking
appears instead to be the one of $SU(2)_L \times SU(2)_R$ into its diagonal
subgroup, the custodial $SU(2)$.
The process under concern can accordingly, now, be also mediated by the
Higgs boson which, because of the Cabibbo-Kobayashi-Maskawa (CKM) rotation,
connects, on one side through the mexican-hat potential, pseudoscalar mesons
with different flavors
\footnote{The occurrence of these flavor changing neutral currents have
already been mentioned in \cite{Machet2} in the decays of the $Z$ boson into
two leptons and two pseudoscalar mesons.}
,
and, on the other side couples through Yukawa couplings to massive neutrinos.

\section{The (non-standard) Higgs contribution to
$\bs{K^+ \rightarrow \pi^+ \ell\bar\ell}$}
\label{section:Hcontrib}

\subsection{Theoretical framework}
\label{subsec:theo}

The theoretical framework has been set in \cite{Machet1} and
\cite{Machet2} (section 2); we work in the approximation where the
electroweak quadruplet of $J=0$ mesons containing the Higgs boson and the three
(pseudoscalar) goldstone bosons is
\footnote{The $N^2/2$ quadruplets of $J=0$ mesons corresponding to $N$ flavors
of quarks are labeled by $(N/2)^2$ real matrices $\mathbb D$ of dimension
$N/2 \times N/2$ and ${\mathbb D}_1$ is the corresponding unit matrix; see
\cite{Machet2} for the notations.}

\begin{equation}
(H, \vec G) = ({\mathbb S}^0, \vec{\mathbb P})({\mathbb D}_1);
\label{eq:HG}
\end{equation}
this is akin to taking as non-vanishing only the flavor-diagonal quark
condensates and to choosing all of them to be identical
\footnote{In this limit, the two neutral kaons are not coupled to the Higgs
boson, which consequently does not participate, in particular, to their
decays into $\pi^0 \ell\bar\ell$.}
.

The quartic term in the ``mexican hat'' potential triggers, after symmetry
breaking, a coupling
between the Higgs boson and two of the three Goldstones of the broken
electroweak symmetry (or, equivalently,
 of the breaking of $SU(2)_L \times SU(2)_R$ into
the custodial $SU(2)_V$ \cite{Machet1}).
Because of the CKM rotation, the Goldstones are
not flavor eigenstates but mixtures of them, which entails new
type of flavour changing neutral currents, connecting in particular
$K^+$ and $\pi^+$ mesons.

On the other side, the Yukawa couplings that are introduced between leptons
and the real quadruplet (complex doublet) (\ref{eq:HG})  to give  Dirac
masses to the former couple the Higgs boson to leptonic mass eigenstates
(which may not be flavour eigenstates in the case of neutrinos).

So, working with two generations ($N=4$), the two vertices involved in the
diagram of Fig.~1 write
\bea
V_{hK\pi} &=& -i\sqrt{2}\,c_\theta s_\theta\, \lambda\, v,\cr
V_{h\ell\bar\ell} &=& i\frac{m^D_\ell}{v/\sqrt{2}},
\label{eq:vertex}
\eea
where $\lambda$ is the coupling constant of the quartic term in the mexican-hat
potential, $c_\theta$ and $s_\theta$ are the cosine and sine of the Cabibbo
angle,
$v/\sqrt{2}$ is the vacuum expectation value of the Higgs boson,
and $m^D_\ell$ is the Dirac mass of the outgoing lepton.
The mass of the Higgs boson is $M_h^2 = \lambda v^2$; as it is supposed to
be much larger than the mass of the incoming mesons, we shall neglect
the momentum dependence of the Higgs propagator, which makes the amplitude
independent of $\lambda$.

\vbox{
\begin{center}
\epsfig{file=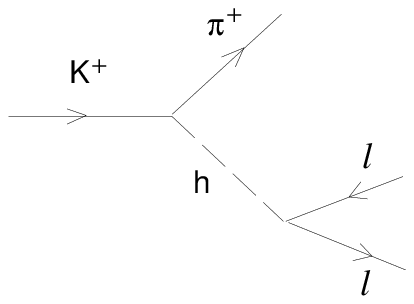,height=5truecm,width=7truecm}
\figskip
{\em Fig.~1: The Higgs mediated decay $K^+ \rightarrow \pi^+ \ell\bar\ell$}
\end{center}
}

\subsection{Calculation of the decay rate for $\bs{K^+ \rightarrow \pi^+
\nu\bar\nu}$}
\label{subsec:calcul}

The calculation includes the normalization factor \cite{Machet2}
\begin{equation}
{\textswab b} = \frac{\la H\ra}{2f_0} = \frac{v}{2\sqrt{2}f_0}
\label{eq:b}
\end{equation}
which relates the fields of dimension $[mass]$ in the Lagrangian to
observed asymptotic mesons; $f_0$ is the generic leptonic decay constant
of pseudoscalar mesons, that we consider to be the same for $K$ and
$\pi$; $\textswab b$ modifies accordingly (see \cite{Machet2}) the phase-space
measure for the outgoing pion, which becomes itself proportional to
${\textswab b}^2$; this  makes the decay rate (\ref{eq:gamma}) below
proportional to $1/{\textswab b}^2$, that is to $f_0^2$,  and
yields a $G_F^3$ dependence though one deals with a tree-level diagram.

The decay rate writes
\begin{equation}
\Gamma_{K^+ \rightarrow \pi^+ \nu\bar\nu} =
(s_\theta c_\theta)^2 \frac{\sqrt{2}f_0^2 G_F^3 (m_\nu^D)^2} {\pi^3 M_K^3}
\left[
\frac{(M_K^2 -M_\pi^2)^3}{3} +
\frac{M_K^2 + M_\pi^2}{2}
   \left( M_K^4 -M_\pi^4 -2M_K^2
                                 M_\pi^2\ln\frac{M_K^2}{M_\pi^2}\right)
\right],
\label{eq:gamma}
\end{equation}

where having neglected their masses in the phase-space integration for neutrinos
enabled to get an analytic expression,
in which the sole dependence on the Dirac neutrino mass $m^D_\nu$ comes from
their coupling to the Higgs boson.

The value of the corresponding branching ratio is plotted on Fig.~2 as a
function of $m_\nu^D$.

\vbox{
\begin{center}
\epsfig{file=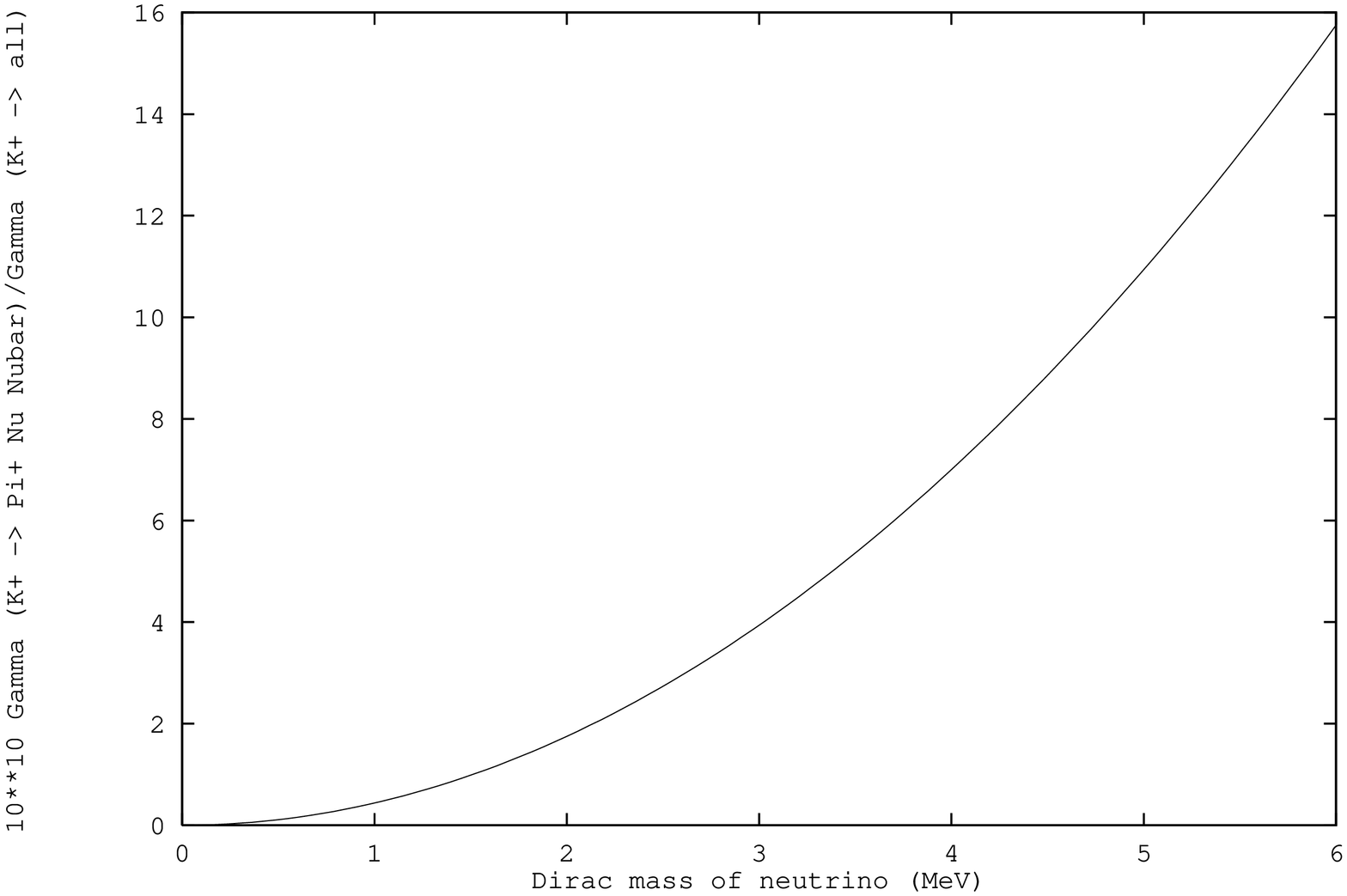,height=14truecm,width=10truecm,angle=-90}
\figskip
{\em Fig.~2: The branching ratio ${\Gamma_{K^+ \rightarrow
 \pi^+ \nu\bar\nu}}/{\Gamma_{K^+\rightarrow all}}$ as a function of
$m^D_\nu$}
\end{center}
}

\subsection{Influence of the neutrino spectrum}
\label{subsec:mixing}

We have supposed a hierarchical scheme for the neutrino masses and only
considered the coupling of the Higgs to the heaviest one.

In case non-sterile neutrinos are roughly degenerate, the three corresponding,
nearly
equivalent,  amplitudes should be added; as a result, Fig.~2 should be read
for $3\,m^D_\nu$ instead of $m^D_\nu$.

As detecting and identifying the outgoing neutrinos by their flavor properties
is well beyond present experimental ability, there is no purpose in
introducing the leptonic mixing matrix and studying a precise channel.

\subsection{Calculation of the decay rates for $\bs{K^+ \rightarrow \pi^+
e^+e^-}$ and $\bs{K^+ \rightarrow \pi^+ \mu^+ \mu^-}$}
\label{subsec:l+l-}

It is important to check that the same mechanism does not grossly alters
the standard predictions in the cases where the two outgoing leptons are
charged (electrons or muons).
The calculations go along the same way, except that one has to keep the
dependence on their masses when performing the phase space integral; the final
evaluation can then only be numerical.

One gets
\begin{equation}
\Gamma_{K^+ \rightarrow \pi^+ e^+ e^-} = 6.4\;10^{-29}\,GeV
\label{eq:gammaEE}
\end{equation}
and
\begin{equation}
\Gamma_{K^+ \rightarrow \pi^+ \mu^+ \mu^-} = 8.37\;10^{-25}\,GeV;
\label{eq:gammaMUMU}
\end{equation}
this corresponds to  the branching ratios
\begin{equation}
BR(K^+ \rightarrow \pi^+ e^+ e^-) = 1.19\;10^{-12}
\label{eq:BREE}
\end{equation}
and
\begin{equation}
BR(K^+ \rightarrow \pi^+ \mu^+ \mu^-) = 1.56\;10^{-8},
\label{eq:BRMUMU}
\end{equation}
to be compared with the experimental values \cite{RPP}
\begin{equation}
BR_{exp}(K^+ \rightarrow \pi^+ e^+ e^-) = 2.74\pm .23\;10^{-7}
\label{eq:BREEexp}
\end{equation}
and
\begin{equation}
BR_{exp}(K^+ \rightarrow \pi^+ \mu^+ \mu^-) = 5\pm 1\;10^{-8}.
\label{eq:BRMUMUexp}
\end{equation}
This shows that the non-standard Higgs contribution is negligeable
in the case of
two outgoing electrons, and within the range of the experimental uncertainty
in the case of two outgoing muons.

In those two last cases, ours is consequently not expected to modify present
theoretical calculations.

\section{An upper bound for the Dirac mass term of the heaviest non-sterile
neutrino}
\label{section:bound}

We suppose that standard calculations could at the maximum account for a
branching ratio\break
\hbox{$BR_{K^+ \rightarrow \pi^+ \nu\bar\nu} \approx 1.5\;10^{-10}$};
having no information on the relative sign between the standard
amplitude and the new Higgs-mediated contribution, 
we can only say that the latter dominates when it yields a partial decay
rate at least twice as large as the limit above,
that is for $m^D_\nu \geq 2.5\,MeV$.

Then the experimental upper bound (\ref{eq:exBRnunu}) entails
\begin{equation}
m^D_\nu \leq 5.5\,MeV
\label{eq:bound}
\end{equation}
which is three times lower than the direct bound \cite{RPP}
\begin{equation}
m_{\nu_\tau} \leq 18.2\, MeV.
\label{eq:directbound}
\end{equation}
The average experimental value (\ref{eq:exBRnunu}) corresponds to
\begin{equation}
m^D_\nu \approx 3\,MeV.
\label{eq:mDaverage}
\end{equation}
This value is much higher than generally presumed order of magnitudes for
masses of non-sterile neutrinos coming form
recent results on solar and atmospheric neutrinos \cite{Bilenky},
setting the scales for the mass splittings,
combined with absolute upper bounds on neutrinos masses \cite{kayserpeccei},
in particular the ones coming from studying the spectrum of the $\beta$
decay of $^3H$ \cite{lobashev}
\begin{equation*}
m_{\nu_e} \leq 3.9\,eV\;(90\%\;CL).
\label{eq:nueabs}
\end{equation*}
The only known mechanism which could account for such a discrepancy between
the observed neutrino mass and a Dirac mass term is the
so-called ``see-saw'' mechanism \cite{see-saw} in which, in addition to the
Dirac mass term, a Majorana mass term ${\cal M} \gg m^D_\nu$ is generated
through a coupling to a new triplet of scalars; it is associated with a new
scale of physics (right-handed gauge fields, grand unified theories
{\it etc}). It is thus of interest to study this case here.
However, as shown below, the only effect of advocating for such a mechanism
is to replace in the expression of the decay rate (\ref{eq:gamma}) the Dirac
neutrino mass $m^D_\nu$ by the one of the lightest Majorana eigenstate.
The conclusion is thus maintained that the new process advocated here has
to be considered only for neutrino masses in the $MeV$ range, and is
negligeable if they lie in the $eV$ range.

After the diagonalisation
of this general mass matrix involving the two types of mass terms
\cite{Bilenky}, the two eigenstates are Majorana neutrinos 
$\nu_1 = i(\nu_L - (\nu_L)^c),\ \nu_2 = \nu_R + (\nu_R)^c$,
with masses respectively
\begin{equation}
m_1 = {(m^D_\nu)^2}/{\cal M},\ m_2 \approx {\cal M}.
\label{eq:seesaw}
\end{equation}
But, while the left-handed neutrino is mostly made of the lightest eigenstate
$\nu_1$, the right-handed one which, in the process under scrutiny,
is coupled through the Higgs boson to the left-handed neutrino, is mostly
made of the heaviest one $\nu_2$, which one does not expect to be produced
here, plus only a very small admixture of the light one, in the proportion 
$m^D_\nu/m_2 \approx m^D_\nu/{\cal M}$.

The decay rate (\ref{eq:gamma}) has thus now to be multiplied by
$(m^D_\nu/{\cal M})^2 \approx (m_1/m^D_\nu)^2$, where we have used
(\ref{eq:seesaw}). This has the global effect of replacing in
(\ref{eq:gamma}) the factor $(m^D_\nu)^2$ by the light neutrino mass
$m_1^2$.
\section{Conclusion}
\label{section:conclusion}

By providing a unified view of $J=0$ mesons which includes the Higgs boson,
the extension \cite{Machet1} of the electroweak standard model enables, in
this sector, predictions which  depart from the Glashow-Salam-Weinberg model.

After $K \rar \pi\pi$ decays \cite{Machet3} and the disintegrations of the
$Z$ boson into two pseudoscalar mesons and two leptons \cite{Machet2}, we
have extended here our investigations to the rare semi-leptonic decays of
kaons; we have shown that, in this framework and unlike in the standard
model, decay rates for $K^+ \rar \pi^+ \nu\bar\nu$ in
agreement with present experimental bounds can be accounted for with
neutrino masses in the $MeV$ range, which is not yet excluded
experimentally.
\vskip 1cm 
\begin{em}
\underline {Acknowledgments}: It is a pleasure to thank S.T. Petcov and
X.Y. Pham for suggestions, comments and advice, and also the referee for
pointing out a misinterpretation in the first version of this work.
\end{em}
\newpage\null
\begin{em}

\end{em}

\end{document}